\documentclass[aps,prb,twocolumn,superscriptaddress,a4paper]{revtex4-1}
\usepackage{amsfonts,amssymb,amsmath}

\usepackage[T1]{fontenc}
\usepackage[utf8]{inputenc}
\usepackage{color}
\usepackage{ifthen}
\usepackage{ulem}

\usepackage[dvips]{graphicx}

\usepackage[hypertex,breaklinks=true,colorlinks=false]{hyperref}
\usepackage[active]{srcltx}

\newcommand{\journal}[4]
{\ifthenelse{\equal{#1}{pr}}{
Phys Rev. \ {\bf #2}, \href{http://link.aps.org/abstract/PR/v#2/e#3}{#3} (#4)}
{\ifthenelse{\equal{#1}{prl}}{
\prl \ {\bf #2}, \href{http://link.aps.org/abstract/PRL/v#2/e#3}{#3} (#4)}
{\ifthenelse{\equal{#1}{prb}}{
\prb \ {\bf #2}, \href{http://link.aps.org/abstract/PRB/v#2/e#3}{#3} (#4)}
{\ifthenelse{\equal{#1}{pra}}{
\pra \ {\bf #2}, \href{http://link.aps.org/abstract/PRA/v#2/e#3}{#3} (#4)}
{\ifthenelse{\equal{#1}{arxiv}}{preprint
\href{http://arxiv.org/abs/#2.#3}{arXiv:#2.#3}}
{\ifthenelse{\equal{#1}{rmp}}{
\rmp \ {\bf #2}, \href{http://link.aps.org/abstract/RMP/v#2/e#3}{#3} (#4)}
{\ifthenelse{\equal{#1}{cond-mat}}{preprint
\href{http://arxiv.org/abs/cond-mat/#2}{cond-mat/#2}}
{\ifthenelse{\equal{#1}{pre}}{
\pre \ {\bf #2}, \href{http://link.aps.org/abstract/PRE/v#2/e#3}{#3} (#4)}
{#1 \ {\bf #2}, #3 (#4)}}}}}}}}}

\newcommand{\journaldoi}[5]{#1\ {\bf #2}, \href{http://dx.doi.org/#5}{#3} (#4)}
\usepackage{tikz}
\usetikzlibrary{arrows}
\usetikzlibrary[shapes.geometric]
\usepackage{leftidx}

\begin{document}
\title{Geometric entanglement of critical XXZ and Ising chains and
Affleck-Ludwig boundary entropies}
\author{Jean-Marie St\'ephan}
\author{Gr\'egoire Misguich}
\affiliation{
Institut de Physique Th\'eorique,
CEA, IPhT, CNRS, URA 2306, F-91191 Gif-sur-Yvette, France.}
\author{Fabien Alet}
\affiliation{Laboratoire de Physique Th\'eorique, IRSAMC, Universit\'e de
Toulouse, CNRS, F-31062 Toulouse, France}

\begin{abstract}
We study the geometrical entanglement of the XXZ chain in its critical regime.
 Recent numerical simulations [Q.-Q. Shi, R. Or\'us, J. O. Fj\ae restad and H.-Q Zhou, New J. Phys. {\bf 12}, 025008 (2010)]
 indicate that it scales linearly with system size, and that the first subleading correction is constant, which was argued to be possibly universal.
 In this work, we confirm the universality of this number, by relating it to the Affleck-Ludwig boundary entropy corresponding
 to a Neumann boundary condition for a free compactified field. We find that  the subleading constant is a simple function of the
 compactification radius, in agreement with the numerics. As a further check, we compute it exactly on the lattice at  the XX point.
 We also discuss the case of the Ising chain in transverse field and show that the geometrical entanglement is related to the Affleck-Ludwig boundary entropy
 associated to a ferromagnetic boundary condition.
\end{abstract}
\maketitle

\section{Introduction}

The geometrical entanglement (GE) is a measure of the multipartite entanglement in a wave-function,
and quantifies the distance to the closest {\it untangled} (separable)
state.\cite{ge2,ge}
Starting from a wave-function $|\Psi\rangle$ for a quantum lattice model with
$N$ sites, one defines a maximal overlap  $\Lambda_{{\rm max}}$ to be
\begin{equation}\label{eq:overlap}
 \Lambda_{\rm max}=\max_{|\Phi\rangle} |\langle \Phi|\Psi\rangle|,
\end{equation}
where the maximization is carried on states $|\Phi\rangle$ which are
 tensor products of single-site states :
\begin{equation*}
 |\Phi\rangle=\bigotimes_{j=1}^{N} |\Phi_j\rangle 
\end{equation*}
The larger $\Lambda_{\rm max}$, the closest it is to a product state, and the
less entangled $|\Psi\rangle$ is.
The GE is then defined as\cite{ge}
\begin{equation*}
 E(|\Psi\rangle)=-\log _2 \Lambda_{\rm max}^2.
\end{equation*}
This quantity allows to quantify the global multipartite entanglement of the
wave function $|\Psi\rangle$, and is useful in the context of quantum computation~\cite{Biham} or state discrimination with local measurements~\cite{Hayashi}.

The GE has also recently gained interest in many-body and condensed-matter physics~\cite{weietal,ge-cm,oruswei,hl10}.
Since it generally scales linearly with the volume ($N$) of the system one also
defines the GE per site:
\begin{equation*}
 \mathcal{E}_N(|\Psi\rangle)=N^{-1} E(|\Psi\rangle).
\end{equation*}
As an example of interesting application, it has been shown that derivatives of
$\mathcal{E}_N$ on large systems can be used
to detect quantum phase transitions.\cite{weietal,oruswei,hl10}.

In this paper we focus on the case of the ground-state $|\Psi\rangle=|G\rangle$ of the Hamiltonian of a 
periodic spin-1/2 XXZ chain:
\begin{equation}\label{eq:H}
\mathcal{H}=-\sum_{j=1}^{L}\left( \sigma_j^x \sigma_{j+1}^x +\sigma_j^y
\sigma_{j+1}^y+\Delta \sigma_{j}^z \sigma_{j+1}^z\right)
\end{equation}
where $L=N$ is the number of sites. We consider the anisotropy parameter in the
range $|\Delta|<1$, so that the system
is gapless (critical). In Ref.~\onlinecite{shietal} it was show numerically that
the GE per spin admits the following asymptotic expansion :
\begin{equation}\label{eq:scaling}
 \mathcal{E}_L(\Delta)=\mathcal{E}_{\infty}(\Delta)+b(\Delta)/L+O(1/L^2)
\end{equation}

The purpose of this paper is to show that the term $b(\Delta)$ appearing in this
expansion can be
 computed analytically in a simple way. Using results of boundary Conformal Field Theory (CFT)~\cite{foo2}, we relate it
 to an Affleck-Ludwig (AL) boundary entropy.\cite{AffleckLudwig}

\section{Relation to the AL boundary entropy}
\label{sec:xxz}

The product state $|\Phi\rangle_{\rm
max}$ that maximizes the overlap in Eq.~\ref{eq:overlap} with the ground-state of Eq.~\ref{eq:H} is
known to be a spin configuration where all spins are parallel \cite{foo1} and 
lie in the $XY$ plane :
\begin{eqnarray*}
 |\Phi\rangle_{\rm max}&=&|\to \to \ldots \to\rangle \label{eq:phi_max}\\
&=&2^{-L/2}\sum_{\{\sigma_j^z=\pm 1\}}\left|\sigma_1^z \sigma_2^z \ldots
\sigma_L^z\right \rangle\\
&=&|{\rm free}\rangle_z
\end{eqnarray*}
Computing $b(\Delta)$ amounts to calculating a subleading contribution to the
scalar product $\leftidx{_z}\langle {\rm free}|G\rangle$.
To do this, we adopt a transfer matrix point of view where $|G\rangle$ 
is interpreted as the dominant eigenstate of the transfer matrix $M$ of some
two-dimensional classical system
(6-vertex like in our case). The classical model is
defined on a cylinder of circumference $L$ and height $L_y\gg L$, with free
boundary conditions for the spins
(or the 6-vertex arrows) at both edges.
In this geometry the free energy $F=-\ln\leftidx{_z\!}\langle {\rm
free}|M^{L_y}|{\rm free}\rangle_z$ can be written\cite{AffleckLudwig} as $F=F_{\rm bulk}+2F_{\rm
boundary}$,
with $F_{\rm bulk} \sim L L_y$ and $F_{\rm boundary}=a L+s+o(1)$. $a$ is the
boundary free energy per unit length, and $s$ is a subleading term in the
boundary free energy.

It is easy to check that $F_{\rm boundary}=-\ln \leftidx{_z\!}\langle {\rm
free}|G\rangle$
in the limit $L_y\gg L$ where only the state $|G\rangle$ contributes.
So, $b(\Delta)$ is simply related to the constant $s$ in the boundary free
energy:
\begin{eqnarray}
 b(\Delta)=-\frac{2}{\ln 2}s.
\label{eq:bs}
\end{eqnarray}

For critical systems $s$ is universal and may therefore be computed in the
continuum limit.
In the XXZ chain (equivalently 6-vertex  case), it corresponds to a compactified free field
with the following (Euclidian) Lagrangian density:
\begin{eqnarray*}
 \mathcal{L}=\frac{1}{8\pi} (\partial_\mu \phi)^2\\
\phi\equiv \phi+2\pi R
\end{eqnarray*}
The compactification radius $R$ is related to the
decay exponents of the correlation functions and to the Luttinger parameter.
In the case of Eq.~\ref{eq:H}, $R$ is a known function of
$\Delta$:\cite{xxzcft}	
\begin{equation}
R(\Delta)=\sqrt{\frac{2}{\pi} \arccos (\Delta)}.
\label{eq:Delta}
\end{equation}
As well-known in boundary CFT\cite{foo2}, free boundary condition for the spins
(or arrows) are encoded in the continuum limit by a Neumann boundary condition
for the free field. So, the boundary
entropy $s$ is in fact the AL\cite{AffleckLudwig} boundary entropy
$s_N$ associated to 
Neumann boundary condition. Its value is known\cite{fsw} and depends solely on the
compactification radius:
\begin{equation*}
s_N=\ln(g_N)\;\;,\;\;g_N=\sqrt{R/2},
\end{equation*}
where $g_N$ is the so-called ``universal non-integer ground-state
degeneracy'',\cite{AffleckLudwig} or ``$g$-factor''.	
Combining this with Eq.~\ref{eq:bs}, our prediction for $b(\Delta)$ is :
\begin{equation}
 b(\Delta)=1-\log_2 R(\Delta).
\label{eq:b}
\end{equation}

\begin{figure}
\begin{center}
\includegraphics[width=6cm,angle=-90]{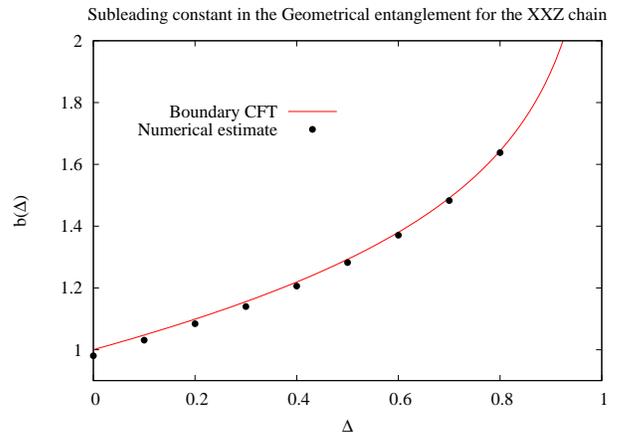}
\caption[1]{(color online) $b(\Delta)$ for the XXZ chain, defined in
Eq.~\ref{eq:scaling}, as a function of the anisotropy parameter $\Delta$.
Symbols are numerical data of Ref.~\onlinecite{shietal} while the solid curve
is the field theoretical calculation (Eqs.~\ref{eq:b} and \ref{eq:Delta}). At
$\Delta=0$, an exact lattice calculation shows that $b=1$.} 
\label{fig:GE}
\end{center}
\end{figure}

As can be seen in Fig.~\ref{fig:GE}, this result matches very well the numerical
data of Ref.~\onlinecite{shietal}. The slight discrepancy
for small $\Delta$ is  likely due to finite-size effects and/or finite-bond
dimension errors in the matrix-product representation of the ground-state.
This can be confirmed by an exact microscopic calculation\cite{foo2} of 
$b(\Delta)$ at the
 free fermion point $\Delta=0$.

The free fermion representation of the chain at $\Delta=0$ allows to obtain\cite{foo3}
$\mathcal{E}_{L}$ on the lattice, as done in Ref.~\onlinecite{weietal}:
\begin{equation}
  \mathcal{E}_{L}=1+\frac{2}{L\ln 2} \sum_{j=1}^{L/4} \ln  \tan
\left(\frac{(2j-1)\pi}{2L}\right).
\end{equation}
From this one gets the expansion in powers of $1/L$ using the Euler-Maclaurin 
formula.\cite{foo4}
The result is:
\begin{eqnarray*}
\mathcal{E}_{L}&=&\mathcal{E}_{\infty}+\frac{b}{L}+O(1/L^2)\\
\mathcal{E}_{\infty}&=&1-\frac{2K}{\pi \ln 2}\simeq 0.158733\\
b&=&1,
\end{eqnarray*}
where $K$ is Catalan's constant. The result can be compared with the numerical
estimations of Ref.~\onlinecite{shietal} :
$\mathcal{E}_{\infty}\simeq 0.1593$ and $b\simeq  0.98$. As expected, the exact lattice calculation confirms the field theory prediction
at $\Delta=0$, namely $b(\Delta=0)=1$ (Eq.~\ref{eq:b}).

It is also worth noticing that the calculation extends to finite magnetizations of the chain,
and simply amounts to move the fermion density away from $1/2$. In
such a case one observes that  $b(\Delta=0)=1$ is independent of the magnetization. This again matches the
field theory result, since, for $\Delta=0$, the compactification radius $R$ is indeed independent of the
magnetization. 

\section{Critical Ising chain}
The GE for the periodic Ising chain in transverse field
at the critical point
has also been considered numerically in Ref.~\onlinecite{shietal}. The
Hamiltonian is now 
\begin{equation}
 \mathcal{H}=-\sum_{j=1}^{L}\sigma_j^x \sigma_{j+1}^x -\sum_{j=1}^{L}\sigma_j^z
\label{eq:Hising}
\end{equation}
and the product state which maximizes the overlap turns out to be a tilted
configuration:\cite{weietal}
\begin{eqnarray}
|\Phi\rangle_{\rm max}&=&\bigotimes_{j=1}^{N}
\left(\cos(\xi/2)|\!\uparrow_j\rangle + \sin(\xi/2)
|\!\downarrow_j\rangle\right)\label{eq:phi_max_ising}\\
\xi&\simeq&0.897101\nonumber
\end{eqnarray}
where $|\!\uparrow_j\rangle$ and $|\!\downarrow_j\rangle$ are the eigenstates of
$\sigma^z_j$.\cite{foo5}
The authors of Ref.~\onlinecite{shietal} found numerically in this case $b\simeq
1.016$.
We will now give an argument based on boundary CFT, similar to that presented in
the case of the  XXZ chain,
which shows that $b=1$.

Again, we wish to interpret the scalar product $\Lambda_{\rm max}$ appearing in
the GE as a boundary contribution to a classical two-dimensional free energy.
When $|G\rangle$ is the ground state of Eq.~\ref{eq:Hising},
the corresponding classical model is a critical two-dimensional Ising model.
If we had to project $|G\rangle$ onto a state where all spins would point in
the $x$ direction (corresponding to an angle $\xi=\pm \pi$) , it would
correspond to a fixed ferromagnetic boundary condition for the Ising model.
On the other hand, if we had to project onto
a state with all the spins pointing in the $z$ direction ($\xi=0$), it would
correspond to a free boundary condition (as
in the case of the XXZ chain). We therefore see that the tilted state of
Eq.~\ref{eq:phi_max_ising}
somewhat corresponds to a
combination of free and fixed boundary conditions for the classical model.
But the  important point is that $\xi\ne0$. For this reason,  $|\Phi\rangle_{\rm
max}$ breaks the $\mathbb Z_2$ symmetry of the model,
which exchanges the $x$ and $-x$ directions.
In such situation, where the boundary condition imposes a non-zero magnetization
at the edge, the long-distance and universal
properties of the boundary
will be equivalent to that of a system with fixed ferromagnetic boundary
condition (all spins pointing in
the $x$ direction).\cite{AffleckLudwig} In other words, the constant term $s$ in the boundary free
energy
will be the same for the tilted boundary condition and for the ferromagnetic
one.
In this case, the AL entropy $s_{\rm fixed}$ is known to be
$-\frac{1}{2}\ln 2$ 
(corresponding to a $g$-factor $g_{\rm
fixed}=1/\sqrt{2}$).\cite{AffleckLudwig,Cardybe}
As conjectured in Ref.~\onlinecite{shietal}, we therefore obtain $b=1$. This argument can also be confirmed
 using the exact formula of Ref.~\onlinecite{weietal} for the GE,
 and then applying the Euler-Maclaurin formula.\cite{foo6}

\section{Relation with R\'enyi entropies for Rokhsar-Kivelson in 2+1 dimensions}
As we have seen in Sec.~\ref{sec:xxz}, the GE for a critical XXZ chain can be
expressed
using a sum of scalar products with all the spin configurations $|i\rangle$ of
the $z$-basis:
\begin{equation*}
 Z=2^{L/2}\Lambda_{\rm max}=\sum_i\langle i | G \rangle.
\end{equation*}
$Z$ can be seen as a partition function and generalized by
introducing a fictitious inverse temperature $\beta>0$:
\cite{foo7} 
\begin{equation}
 Z(\beta)=\sum_{i}\big(\langle i|G\rangle\big)^\beta,
\label{eq:Z}
\end{equation}
It turns out that $Z(\beta)$ is related to
R\'enyi entanglement entropies of a semi-infinite cylinder in some
two-dimensional Rokhsar-Kivelson wave functions, as studied in
Refs.~\onlinecite{stephan09,stephan10} (the R\'enyi parameter being $n=\beta/2$).
Using numerical and field-theoretical approaches, it was found that $\ln
Z(\beta)$ scales as $ \ln Z(\beta)=aL+\gamma$, where the constant term $\gamma$ is universal and given by 
\begin{equation}
\label{eq:gamma}
 \gamma=(1-\beta/2)\ln R +\frac{1}{2}\ln (\beta/2).
\end{equation}
So, for the XXZ chain, the GE
appears to be a special case $\beta=1$ of this partition function.

We finally comment on the the limit $\beta\to\infty$, where only the spin
configuration $|i\rangle_{\rm max}$
of the $z$ basis which have the largest weight in $|G\rangle$ contributes to
$Z(\beta)$.
For the present XXZ chain,
$|i\rangle_{\rm max}$ is the ferromagnetic state $|i\rangle_{\rm
max}=|\uparrow\cdots\uparrow\rangle$.\cite{stephan09}
Taking $\beta\to\infty$ in Eq.~\ref{eq:Z} and Eq.~\ref{eq:gamma} we get $\ln \langle
\uparrow \cdots \uparrow | G\rangle = -\frac{1}{2}\ln R$
(omitting the extensive term). So, we see that the scalar product of
$|G\rangle$ with both {\it (i)} the uniform configuration where spins point in the $z$
direction and {\it (ii)} the configuration where spins point in the $x$ direction, have very similar subleading terms related to the
logarithm of the compactification radius.
In the CFT language, the former scalar product (appearing in $Z(\beta=\infty)$)
corresponds to a Dirichlet boundary condition for the free field, as opposed to
Neumann for the scalar product appearing in the GE and $Z(\beta=1)$.

\section{Conclusion}

In conclusion, we have related subleading terms of the geometrical entanglement of the critical XXZ and Ising spin chains to their AL boundary entropies,
 showing that this correction is universal. Whether such universal behaviour in the scaling
 of the geometrical entanglement (or of the scalar product of the ground-state with a given separable state in general)
 can be found for other physical systems is an open issue that is worth pursuing. 
	
\section{Acknowledgments} 

We wish to thank J\'erome Dubail and Vincent Pasquier for stimulating discussions. This work is supported by the French ANR program ANR-08-JCJC-0056-01 (FA).


\begin{thebibliography}{99}
\bibitem{foo1}
This was demonstrated numerically in Ref.~\onlinecite{oruswei}. Notice however that in this reference the antiferromagnetic chain was considered. Therefore, the result in this case is : $|\Phi\rangle_{\rm max}=|\rightarrow\leftarrow \rightarrow \leftarrow \ldots\rightarrow \leftarrow\rangle$.
\bibitem{foo3}
For simplicity we only consider the case $L \equiv 0 \mod 4$. Notice that using Eq.~\ref{eq:Z}, this result can also be interpreted (and recovered) as the partition function of a discretized gas of particles on an annulus interacting trough a 2D Coulomb potential (discretized Dyson gas) at inverse temperature $\beta=1$. See Ref.~\onlinecite{stephan09,gaudin} for more details.
\bibitem{foo4}
$b$ can easily be deduced from the constant in the asymptotic expansion of $\sum_{j=1}^m \ln \tan \left[(j-1/2)\pi/(8m)\right]$. Then, a way to proceed is to cut the sum in two:  $\sum_{j=1}^m=\sum_{j=1}^{\sqrt{m}}+\sum_{j=\sqrt{m}+1}^m$. In the first sum, the tangent can be linearized so as to apply Stirling's
formula. The asymptotics of the second sum  can be accessed using the Euler-Maclaurin formula. The dominant term $\mathcal{E}_\infty$ has been previously computed in Ref.~\onlinecite{weietal}.
\bibitem{foo5}
Introducing $f(\xi)=\sqrt{3\cos^2(\xi/2)-2}$, the optimal angle $\xi$ is in the thermodynamic limit solution of 
\begin{eqnarray}\nonumber
\cos (\xi/2)f(\xi)-\frac{2}{\pi}\textrm{arctanh} \left(\frac{f(\xi)}{\cos(\xi/2)}\right)=0.
\end{eqnarray}
\bibitem{foo6}
For example, when $L$ is even :
\begin{eqnarray}\nonumber
\langle \Phi | G\rangle=\prod_k \left[\sin^2 (\xi/2)\sin (k/4)+\cos^2(\xi/2)\cot (k/2) \cos(k/4)\right]
\end{eqnarray}
where $k=(2m+1)\pi/L$, $m\in \{0,\ldots,L/2-1\}$. The subleading constant in the asymptotic expansion of 
$-2 \log_2 \langle \Phi|G\rangle$ can be shown to be 
\begin{equation}\nonumber
 b(\xi)=\left\{\begin{array}{ccc}0&,&\xi= 0 \mod 2\pi\\1&,&\textrm{otherwise}\end{array}\right.
\end{equation}
In the boundary CFT language, $\xi=0$ corresponds to the free boundary state ($s_{\textrm{free}}=0$)
 which is unstable under the renormalization group flow.
\bibitem{foo7}
With our (ferromagnetic) convention for the XY part of the interactions in the XXZ Hamiltonian, all the $\langle i | G\rangle$ are positive.
\end{thebibliography}
\end{document}